\documentclass[aip,apl,reprint]{revtex4-1}
\usepackage{graphicx}
\usepackage{siunitx}
\usepackage{natbib}
\usepackage{amsmath}

\draft % marks overfull lines with a black rule on the right

\begin{document}

%\preprint{}

\title{A split-cavity design for the incorporation of a DC bias in a 3D microwave cavity} %Title of paper



\author{Martijn A. Cohen}
\author{Mingyun Yuan}
\author{Bas W. A. de Jong}
\author{Ewout Beukers}
\author{Sal J. Bosman}
\author{Gary A. Steele}
%\thanks{}
\affiliation{Kavli Institute of Nanoscience, Delft University of Technology, 2628 CJ Delft, The Netherlands}


\date{\today}

\begin{abstract}
We report on a technique for applying a DC bias in a 3D microwave cavity. We achieve this by isolating the two halves of the cavity with a dielectric and directly using them as DC electrodes. As a proof of concept, we embed a variable capacitance diode in the cavity and tune the resonant frequency with a DC voltage, demonstrating the incorporation of a DC bias into the 3D cavity with no measurable change in its quality factor at room temperature. We also characterize the architecture at millikelvin temperatures and show that the split cavity design maintains a quality factor $Q_\text{i} \sim \num{8.8e5}$, making it promising for future quantum applications.
\end{abstract}


\maketitle %\maketitle must follow title, authors, abstract and \pacs

In recent years, interest in 3D waveguide cavity resonators has been revived in the Josephson junction quantum bit (qubit) community. \cite{paik2011observation,  kirchmair2013observation, novikov2013autler, riste2013deterministic, sun2014tracking, flurin2015superconducting, reagor2016quantum} Significantly enhanced relaxation and decoherence times on the order of hundreds of microseconds have been demonstrated for qubits in 3D cavities.\cite{reagor2016quantum} The 3D architecture has also enabled the measurement of entangled qubits, \cite{flurin2015superconducting} single-photon Kerr effect \cite{kirchmair2013observation} as well as the Autler-Townes effect. \cite{novikov2013autler} So far, these results are achieved without the need of applying DC currents or voltages to the devices embedded in the cavities. However, the ability to do so would significantly expand the application of 3D cavities, just as the introduction of a DC bias into a coplanar waveguide cavity made it much more versatile. \cite{chen2011introduction, bosman2015broadband, de2014galvanically} For example, such a DC bias could enable the tuning of a membrane in a 3D optomechanical device. \cite{yuan2015large} For qubit measurements, an on-chip flux bias can be added by applying a DC current bias to allow frequency tunability of the qubit.

\begin{figure}
	\includegraphics{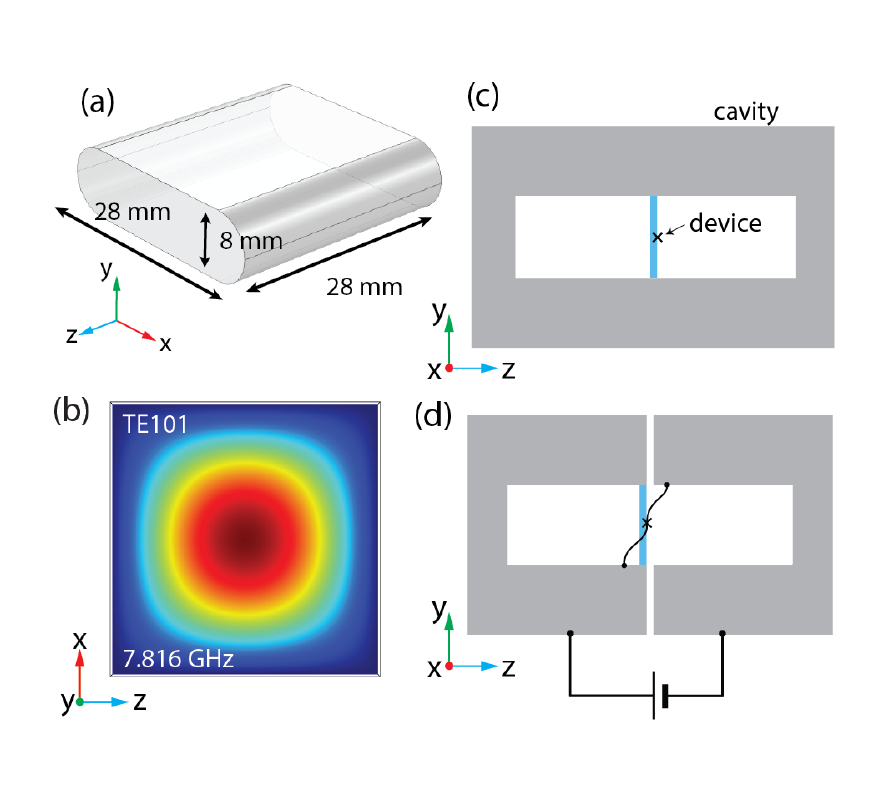}
	\caption{\label{fig1} Implementation of a galvanically accessible 3D cavity by a split-cavity design. (a) Design of the 3D cavity. (b) Electric field magnitude of TE101 mode modelled by COMSOL. (c) A standard implementation of a nanodevice embedded inside 3D cavity without galvanic access. The blue line represents the substrate on which the device is fabricated, silicon or sapphire, for example. (d) Our proposed modification to the cavity geometry to allow a DC bias to be applied. }
\end{figure}

Having galvanic access to a quantum device inside a 3D cavity presents a significant challenge. Planar geometries have a 1D confinement of the electromagnetic field and thus allows access from the side of the cavity. \cite{chen2011introduction, bosman2015broadband} In 3D architectures currents run along a closed 3D surface: incorporating wires into devices inside the cavity without compromising the cavity quality factor is challenging. One study reports a technique of doing so by inserting a sample with wires on a PCB board into a hole in the side of the 3D cavity. \cite{kong2015introduction} These DC wires enabled manipulation of a quantum dot inside the cavity, although this method suffered from a relatively low quality factor $Q_\text{L} = 1350$.

\begin{figure}
	\includegraphics{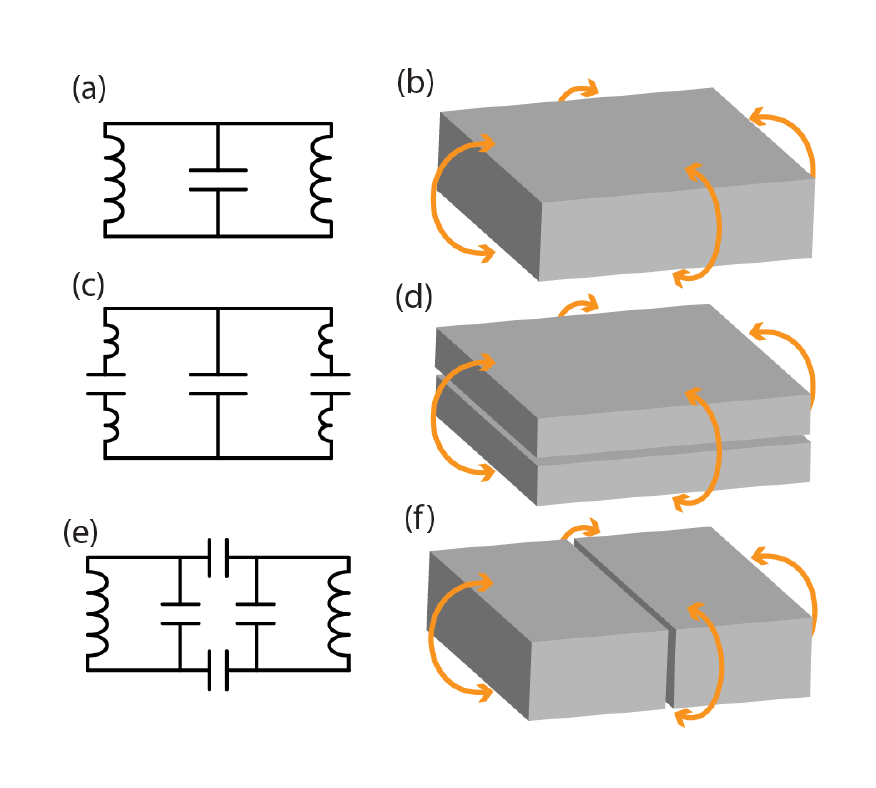}
	\caption{\label{fig2} Minimizing losses of TE101 mode of a split-cavity through choice of orientation of splitting the cavity. (a) An LC circuit representation of a cavity resonating in the fundamental mode. (b) A corresponding 3D representation of a box cavity where the orange arrows show the direction of current flow. For the TE101 mode, the current paths along the side walls act as inductors, and the top and bottom walls as capacitor plates. (c-d) A circuit diagram and 3D representation of one proposed method of galvanically separating the box cavity. When introducing a separation perpendicular to the current flow, this adds capacitances between the inductors of the fundamental mode. (e-f) A circuit diagram and 3D representation of the method of separating the box cavity we use here for the incorporation of the DC bias. The plane of separation is parallel to the direction of current flow. }
\end{figure}

Here, we introduce a method for biasing a 3D cavity while maintaining an internal quality factor of $Q_\text{i} = 8.75 \pm 0.03 \times 10^5$. The dimensions of our near-rectangular cavity are illustrated in Fig.~\ref{fig1}(a). The rounded corners in the xy-plane have a diameter equal to the height of the y-coordinate. According to a finite element simulation shown in Fig.~\ref{fig1}(b) the resonant frequency of the lowest TE101 mode is expected to be around 7.816~GHz. The color corresponds to the magnitude of the electric field in the zx-plane, at the center of which the field reaches maximum. The measured resonant frequency of the bare cavity is $f_0=7.743$~GHz at room temperature and ambient conditions. The loaded Q-factor at room temperature is $Q_\text{L}=3920$. Many implementations use a 3D cavity that is cut in half to allow for a nanodevice to be placed inside. The device is usually fabricated on a dielectric substrate and placed in the center to allow maximum interaction with the electric field. The idea in this article is to use this separation to our advantage by keeping the two halves galvanically separated and applying a DC voltage across it to bias the nanodevice. Such a connection could be achieved through a custom-designed substrate including connections to the back side of the substrate.

\begin{figure}
	\includegraphics{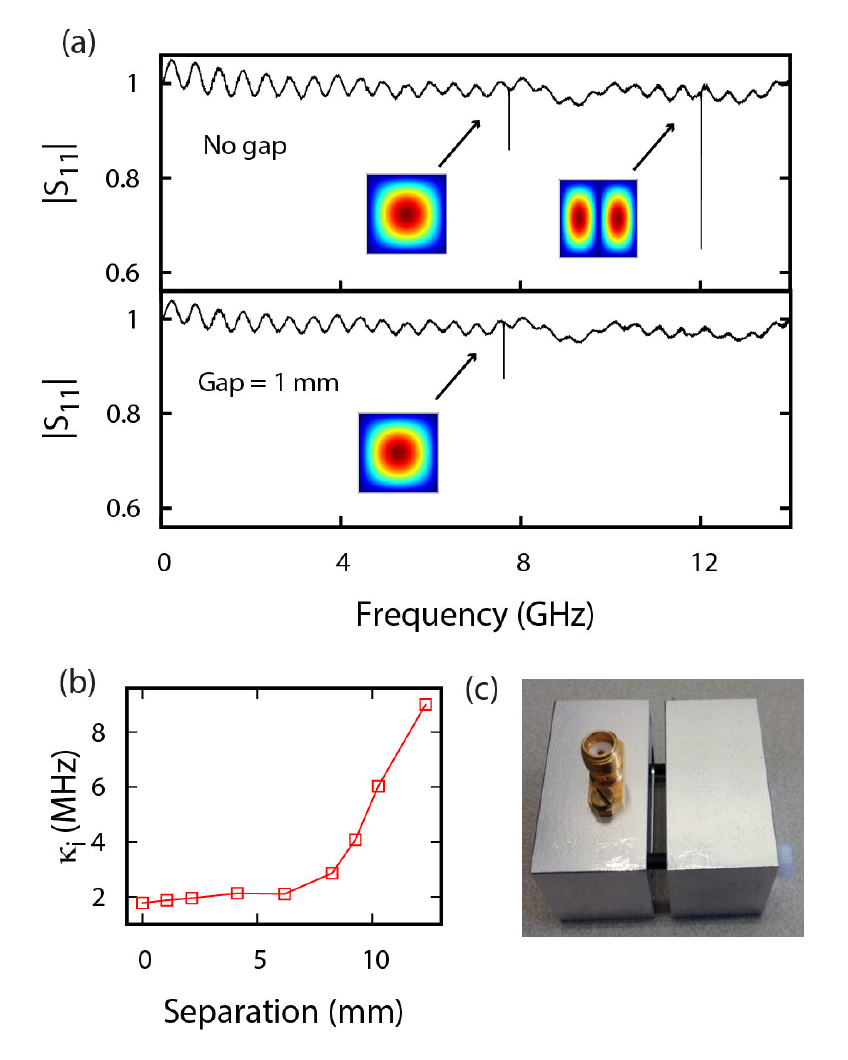}
	\caption{\label{fig3} Experimental observations from room temperature measurements showing that TE101 mode is not affected to first order even for relatively large gaps. (a) Full spectrum of the reflection coefficient for a (top) closed cavity, and (bottom) a cavity separated by 1.07~mm measured at room temperature. The disappearance of the second mode is attributed to the restriction of current across the two halves. (b) A graph of the internal dissipation rate $\kappa_\text{i}$ as a function of cavity separation. (c) A picture of the setup we used to separate the cavity halves. The screws used to connect them were surrounded by an insulating layer. }
\end{figure}

In designing the location of the split in the rectangular 3D cavity, it is important to consider how currents flow within such a resonant structure. For the fundamental mode of the cavity, electrons oscillate back and forth between the plates with the largest area via the side walls. One can see this as a simple LC circuit, where the top and bottom plates act as a capacitor, and current path along the side walls as an inductance. There are two ways we can split this cavity, parallel or perpendicular to the direction of current flow. Fig.~\ref{fig2}(c-d) shows a split cavity where the cut is perpendicular. Such a geometry creates additional capacitances in between the inductors which turns it into a dipole antenna, potentially causing losses by radiation. An alternative geometry would be to split the cavity as depicted in Fig.~\ref{fig2}(e-f). The circuit diagram changes by splitting the capacitance of the LC circuit and separating these by two more capacitances. Since there is a far lower current density at the split, there likely is less chance of radiation losses. A thorough investigation into the losses from seams in cavities is given in Ref.~\onlinecite{brecht2015demonstration}.

\begin{figure*}
	\includegraphics{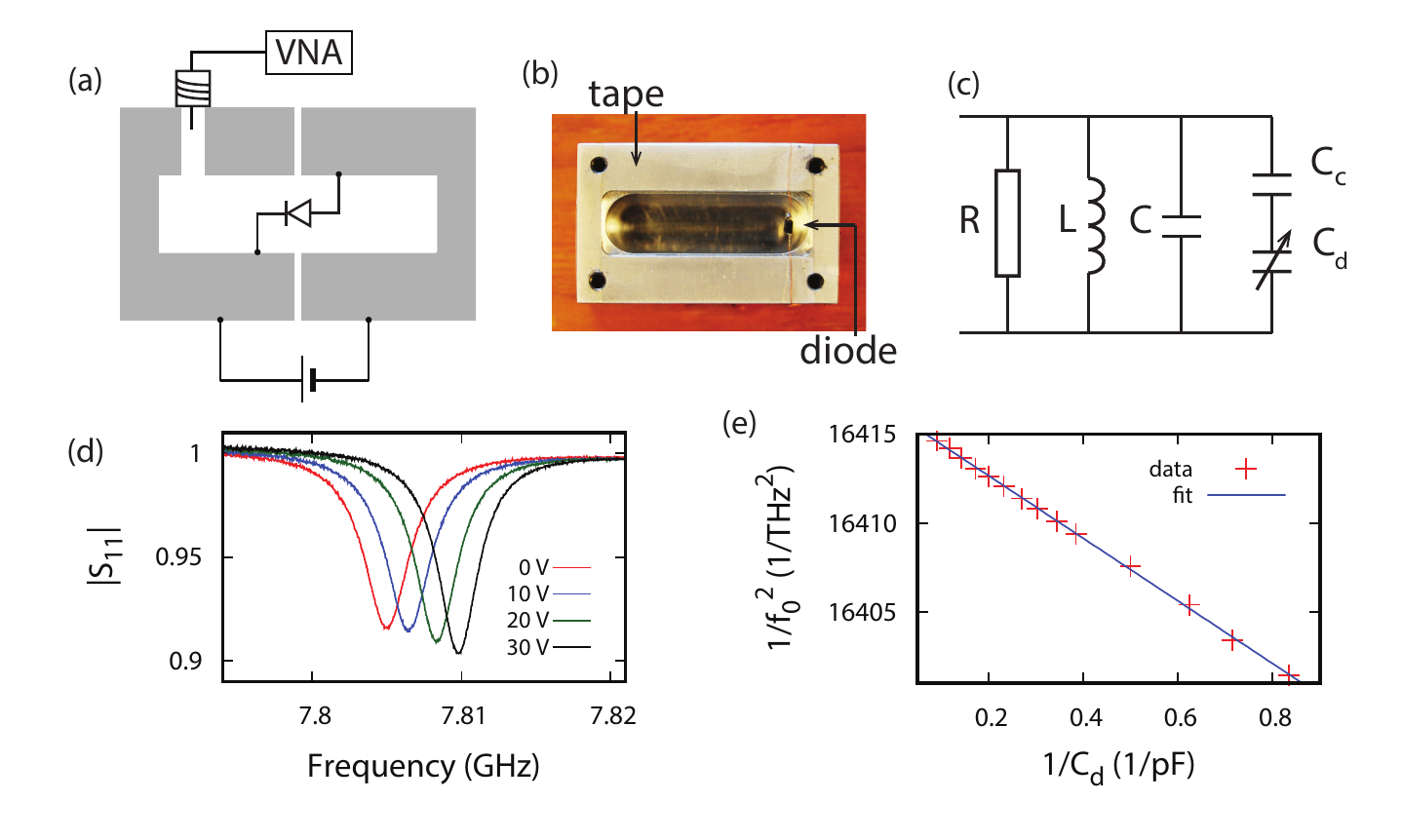}
	\caption{\label{fig4} Proof-of-concept demonstration of a DC tuneable 3D cavity using an embedded reverse biased diode. (a) Schematic of the experimental setup. The diode is reverse-biased by an external voltage applied across the cavity and the resonant frequency is measured reflectively. (b) A photograph of one half of the aluminum cavity. (c) An equivalent circuit diagram. The tuneable diode capacitance $C_\text{d}$ is coupled in parallel to the main cavity through $C_\text{c}$. (d) Representative $S_{11}$ curves measured by the vector network analyzer at room temperature as the bias voltage is varied. The resonant frequency is shifted but the quality factor preserves. (e) Inverse square of the resonant frequency as a function of the inverse of the diode capacitance. }
\end{figure*}

Similar to the coaxial pin used to couple microwaves in and out of the cavity via the readout port, adding a DC wire that fully enters into the cavity would have a significant impact on the cavity Q-factor as it would act as an antenna and contribute heavily to radiative losses. It is in principle possible to partially suppress this leakage using low pass filters, although this presents challenges in terms of the loss rates of the elements and maintaining sufficient grounding to limit radiative losses. In the approach presented here, we avoid such potential problems by separating the two hemispheres of the cavity with a thin layer of insulator and using them directly as electrodes. The interface of the two halves forms a large capacitor that acts as a low-pass filter. We directly apply the DC voltage on the aluminum blocks, which can then be accessed from inside the cavity to connect the DC signals to the device, as illustrated in Fig.\ref{fig1}(d). Using superconducting wires from the two halves of the cavity to devices embedded in it, such as qubits, one can then apply voltage or current biases to the devices. This could also potentially be implemented with traces directly on the sapphire substrates used to hold the devices in the cavity. The bandwidth of this DC bias scheme is limited by the capacitance between the two hemispheres and can be carefully engineered. For a 50~$\Omega$ source impedance supplying the DC signals and a 100~$\mu$m layer Kapton tape insulator with a 180~pF capacitance between the hemispheres, we estimate the bandwidth for the DC signals to be around 18~MHz, and could be even higher if the driving were provided by a lower impedance source. 

To demonstrate the viability of biasing a split 3D cavity, we first present measurements studying how the quality factor is affected by introducing this split at room temperature. In Fig.~\ref{fig3}(a) we show a full spectrum sweep as allowed by our vector network analyzer when the cavity is closed and when the cavity halves are separated by around 1.07~mm. Both configurations display a fundamental frequency, however with a slight shift: 7.74~GHz when closed, 7.62~GHz when opened, about a 1.6\% change. This percentage shift was predicted well by finite element simulations by increasing the length of the box resonator, ignoring the open walls in the region of the gap. There is also a second mode, TE201, at 12.02~GHz, but this can only be detected when the cavity is closed. One can understand this by the fact that the TE201 mode has a large component of current that flows across the split. When the two halves are separated, this current is presented with a large impedance from the capacitor and is therefore strongly suppressed. Even up to very large gaps on the order of millimeters, the linewidth of the TE101 mode, shown in Fig.~\ref{fig3}(b), is relatively constant, indicating that at room temperature, the radiative losses are negligible even for millimeter gaps between the two halves. 

\begin{figure}
	\includegraphics{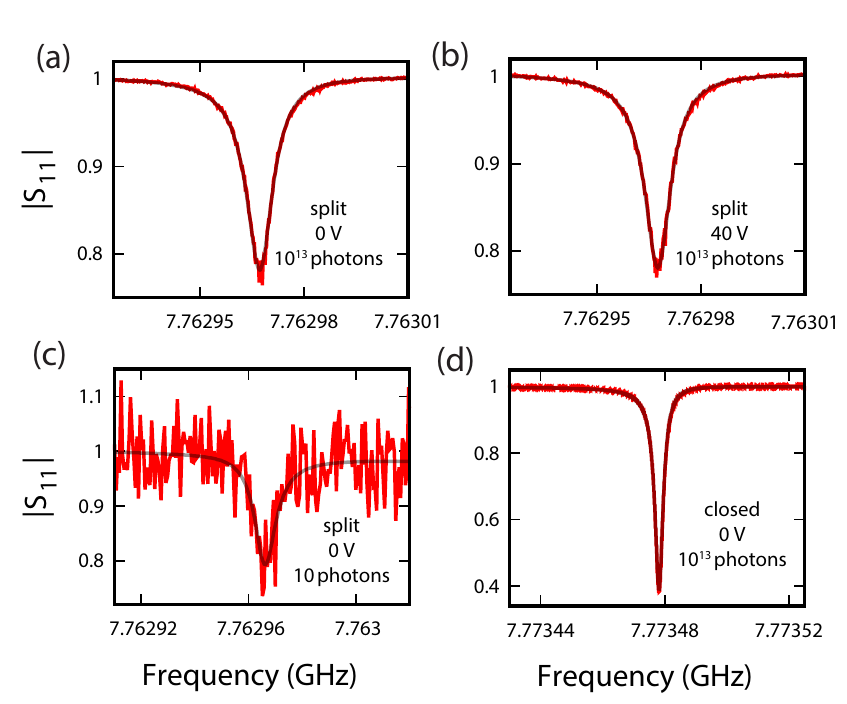}
	\caption{\label{fig5} 
		Characterization of the internal losses of a split cavity at 16~mK inside a dilution refrigerator. Reflection coefficient measurement of a split cavity separated by 100~$\mu$m Kapton tape at $P_\text{in} = -30$~dBm and a DC voltage of 0~V (a) and 40~V (b) with a quality factor $Q_\text{i} = 8.75 \pm 0.03 \times 10^5$ and $Q_\text{i} = 8.82 \pm 0.03 \times 10^5$, respectively. (c) A reflection measurement at $P_\text{in} = -148$~dBm (average photon occupation $N = 10$), with $Q_\text{i} = 8 \pm 1 \times 10^5$. (d) Reflection measurement of a closed cavity at $P_\text{in} = -30$~dBm showing an intrinsic quality factor $Q_\text{i} = 2.748 \pm 0.001 \times 10^6$. Black transparent lines indicate fits to the data. }
\end{figure}

To demonstrate the principle of DC operation of a device inside a cavity, we embed a diode (NXP BB131) inside the cavity. The diode capacitance can be tuned by a reverse voltage bias and the cavity resonant frequency is subsequently shifted. The interface of the left and right halves are separated by a thin layer of scotch tape which acts as an insulator (0.05~mm thickness), and nylon screws are used to avoid shorts. Inside, the anode (cathode) of the diode is attached to the right (left) half of the cavity. An external DC voltage source is attached to the aluminum blocks, reverse-biasing the diode via the cavity walls. One half of the actual cavity is shown in Fig.~\ref{fig4}(b). An equivalent parallel circuit model is illustrated in Fig.~\ref{fig4}(c), in which the resistance $R$, the inductance $L$ and the capacitance $C$ model the basic resonance while the capacitance of the diode $C_\text{d}$ is coupled to the circuit in the model via an effective coupling capacitor $C_\text{c}$.

We perform a reflection measurement with an input power of 0~dBm. In  Fig.~\ref{fig4}(d), we plot the reflection coefficient $S_{11}$, normalized by the off-resonant background as a function of the frequency $f$. The resonant frequency $f_0$ shifts to higher values as we increase the reverse bias voltage $V$ applied on the diode. The quality factors of the cavity show no signs of decrease for $V$ as high as 30~V, and also show no systematic change compared to the quality factor of the cavity without diode at room temperature. The loaded quality factor is  $Q_\text{L}\approx2084$ at 0~V and $Q_\text{L}\approx2292$ at 30~V. From the circuit diagram in  Fig.~\ref{fig4}(c) we can derive $\omega_0^{-2}=L(C+(1/C_\text{c}+1/C_\text{d})^{-1})$, where $\omega_0=2\pi f_0$. Assuming $C_\text{d}\gg C_\text{c}$, rewriting $1/\omega_0^2$ as $y$ and $1/C_\text{d}$ as $x$, the above equation can be rearranged into a linear equation $y=L(C+C_\text{c})-LC_\text{c}^2x$. Using the value of $C_\text{d}$ corresponding to each $V$ provided by the diode data sheet, we plot $1/f_0^2$ as a function of $1/C_\text{d}$ and perform a linear fit as shown in Fig.~\ref{fig4}(e). We further approximate the value of the characteristic impedance $Z_0=\sqrt{L/C}$ to be $Z_\textrm{TE101}\approx540$~$\Omega$, and obtain the following parameters: $L\approx10$~nH, $C\approx35$~fF, $C_c\approx6.65$~fF.

To investigate the feasibility of the split-cavity technique for quantum experiments, in Fig.~\ref{fig5} we show the response of a split-cavity design made from 6061 aluminum alloy measured at a temperature of 16~mK. Kapton tape of 100~$\mu$m thick was used to galvanically isolate the two halves of the cavity. Fig.~\ref{fig5}(a) shows the reflection coefficient without any voltage bias, and Fig.~\ref{fig5}(b) shows the same when applying 40~V. Both measurements reveal similar quality factors, $Q_\text{i} = 8.75 \pm 0.03 \times 10^5$ and $Q_\text{i} = 8.82 \pm 0.03 \times 10^5$, respectively, where the errors are statistical errors from the fit. Although the internal quality factor from such fits, determined by the depth of the dip in the reflection coefficient, can be include systematic error from cable reflections, we also estimate a lower bound of the internal quality factor from the loaded quality factor obtained from the resonance linewidth. For all cryogenic measurements reported here, we find a lower bound of $Q_\text{i} > 7.5 \times 10^5$ (see SI for discussion).  Fig.~\ref{fig5}(c) shows a measurement at a cavity input power $P_\text{in} = -148$~dBm (10 photons) with a fit that corresponds to an intrinsic quality factor of $Q_\text{i} = 8 \pm 1 \times 10^5$. This independence of internal losses to input power has been observed in previous experiments and is one of the key advantages of 3D cavities over planar resonators.\cite{reagor2013reaching} For the closed cavity with no Kapton tape, we find an intrinsic quality factor $Q_\text{i} = 2.748 \pm 0.001 \times 10^6$. Although the split cavity has a smaller quality factor, the reported $Q_\text{i} > 7.5 \times 10^5$ is sufficient to enable sensitive quantum experiments in a 3D microwave architecture.

In conclusion, we have demonstrated a simple method of applying a DC bias up to 40~V in a 3D microwave cavity that maintains a high quality factor. We show that the intrinsic quality factor can be as high as $Q_\text{i} = 8.75 \pm 0.03 \times 10^5$ at dilution refrigerator temperatures, making it suitable for use in 3D implementations of microwave nano and quantum devices. \\

\textit{Supplementary Material} See supplementary material for a thorough documentation of the cryogenic measurements. \\

\textit{Acknowledgements} We would like to thank V. Singh for support and discussion. This work is funded by the Netherlands Organization for Scientific Research (NWO), the Dutch Foundation for Fundamental Research on Matter (FOM), and the European Research Council (ERC).

\bibliography{3d_split}

\onecolumngrid
\pagebreak
\renewcommand{\theequation}{S\arabic{equation}}
\renewcommand{\thefigure}{S\arabic{figure}}
\renewcommand{\thetable}{S\arabic{table}}

\renewcommand{\thesection}{S\arabic{section}}
\renewcommand{\thesubsection}{S\arabic{subsection}}
\renewcommand{\thesubsubsection}{S\arabic{subsubsection}}
\renewcommand{\bibnumfmt}[1]{[S#1]}
\renewcommand{\citenumfont}[1]{S#1}
\setcounter{equation}{0}
\setcounter{figure}{0}
\setcounter{table}{0}

\begin{center}
	\textbf{\large Supplementary Material: A split-cavity design for the incorporation of a DC bias in a 3D microwave cavity}
\end{center}

\section{Cryogenic measurements}

Cryogenic  measurements were done inside of a Triton dilution refrigerator. A picture showing the loading chamber containing the split-cavity is shown in Fig.~\ref{fig1}. The cavity was split by Kapton tape which offers electrical insulation but relatively high thermal conductivity. The half with the SMA connection for the reflection measurement was electrically grounded and thermally anchored with a copper block attached to the bottom plate. The other cavity half was floating and attached to the center pin of another coax which supplied the DC bias. The coaxial cables were made of gold-plated copper-beryllium. Before cooling down the setup, we first test the conductive paths of the cables using a multimeter to make sure everything is sufficiently galvanically connected.

Fig.~\ref{fig2} shows a graph of all the DC measurements taken. From 0 to 40 V the quality factor has a flat trend, which demonstrates the independence of DC voltage to internal losses of the split cavity.

A second measurement sweep was performed in which the input power to the cavity was modulated. Fig.~\ref{fig3} shows two of such data points, along with their fits. The fact that these data points can almost perfectly be overlayed shows that there equally is no dependence on input power to the internal losses of the cavity design.  

\section{Fitting procedure}

We start by unwrapping the phase, which corresponds to eliminating the electrical length of the cables running to the inside of the dilution refrigerator. The second step is to rotate the quadratures so that the real part purely represents the dissipative component and the imaginary part purely represents the dispersive component of the response. Along with this rotation the response is scaled such that off-resonance the reflection is unity. Lastly, the response is fitted using a least-squares method with the following formula:

$$
S_{11}(\omega) = \frac{2\kappa_{e}e^{i\theta}}{\kappa + 2i(\omega-\omega_0)} - 1
$$

The exponential factor in the numerator approximates the asymmetric lineshape one would see when parasitic reflections interfere significantly (i.e. a Fano lineshape). In the case of very large cable resonance and strongly overcoupled cavities, there can be offsets of the quadratures that we cannot systematically eliminate from the fit, leaving systematic error in the extraction of $Q_i$ from the data. On the other hand, from the rotated quadratures plotted in figure S4, one can reliable extract the loaded quality factor $Q_L$ from the resonance linewidth. In this case, we can then place a conservative lower bound on $Q_i$ given by $Q_L$.Table S1 states all the components for the quality factors for the split and closed cavities measured at cryogenic temperatures, along with Fig.~\ref{fig4} showing the fits in quadrature format.

\begin{figure}
	\includegraphics{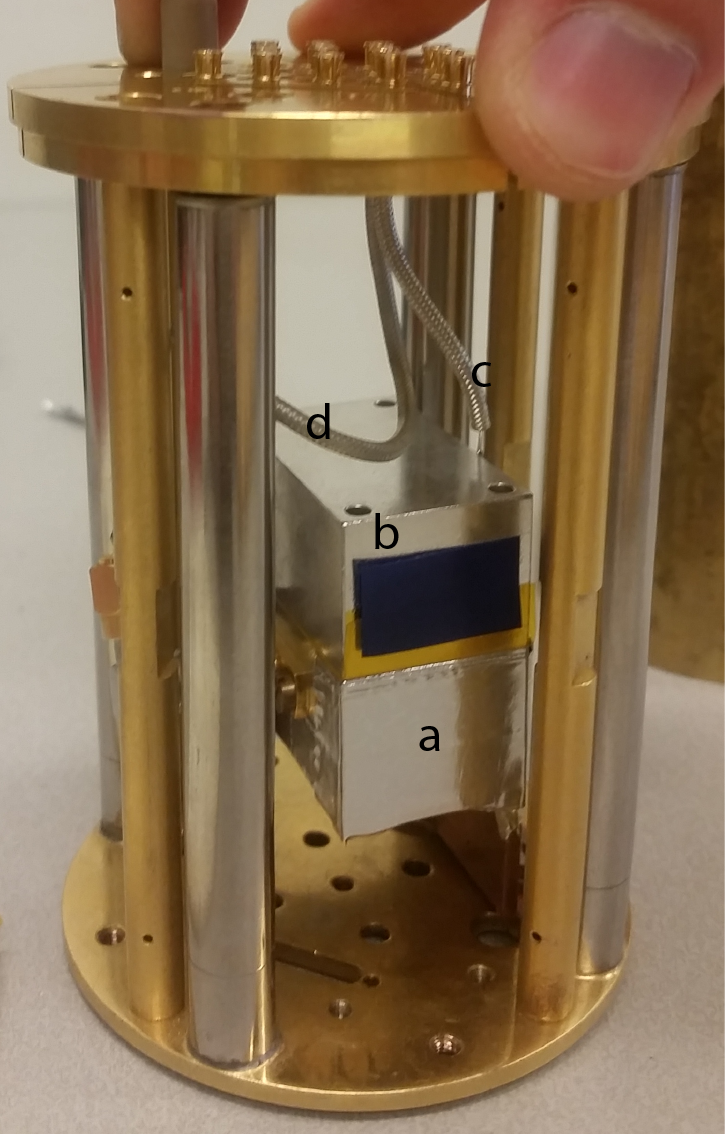}
	\caption{\label{fig1} Photograph of wiring setup while performing cryogenic characterization of the split-cavity design. Label (a) shows the bottom half of the split cavity design, which is grounded and thermally anchored by a copper block in the background. Label (b) shows the top half, which is completely floating, except for a galvanic connection to the center pin of a coax as shown by label (c). The reflection measurement is done through wire labelled (d) which connects to the bottom half of the cavity with an SMA connector. }
\end{figure}

\begin{figure}
	\includegraphics{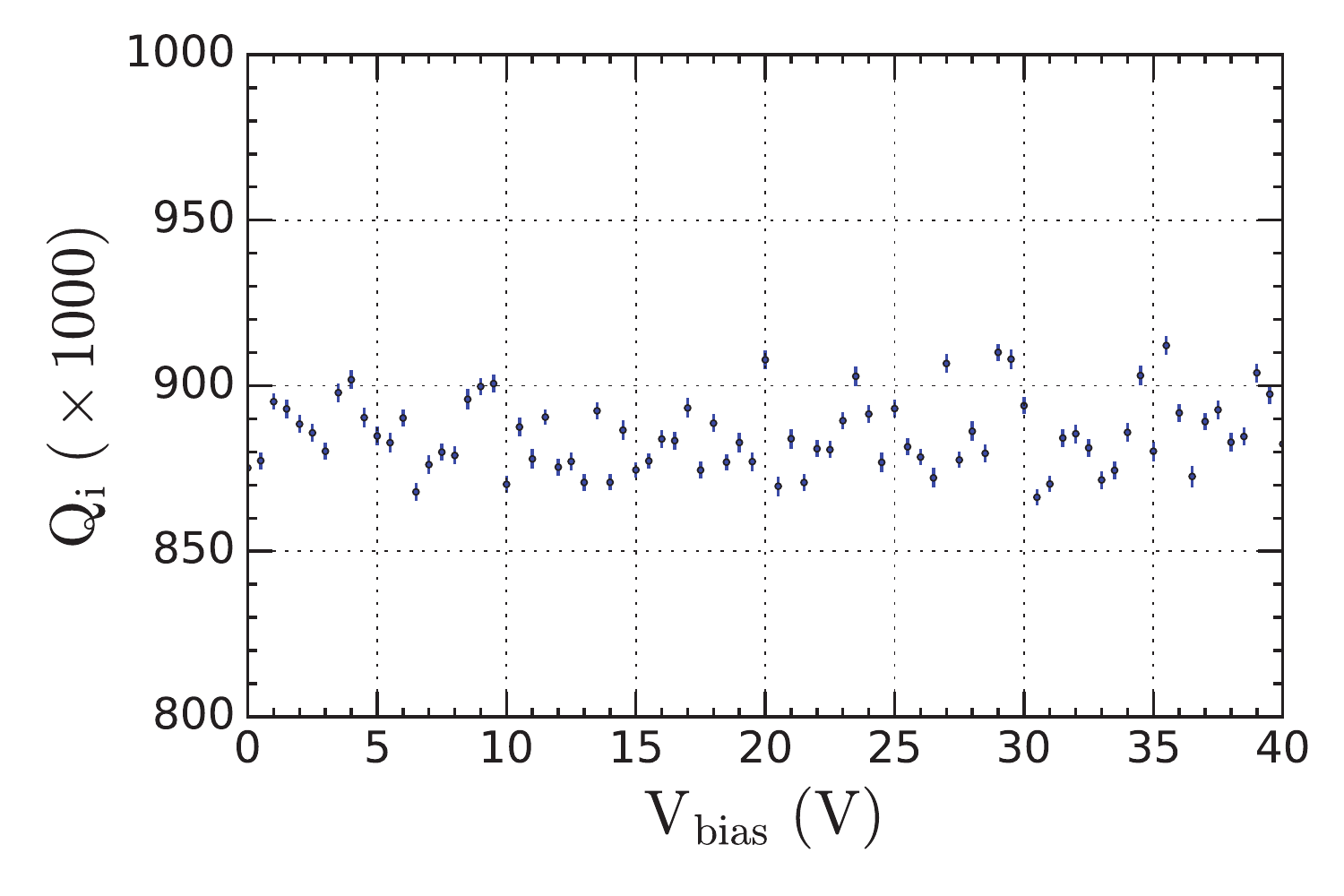}
	\caption{\label{fig2}  Graph of all intrinsic quality factor measurements as a function of bias voltage. Error bars indicate the fitting error. }
\end{figure}

\begin{figure}
	\includegraphics{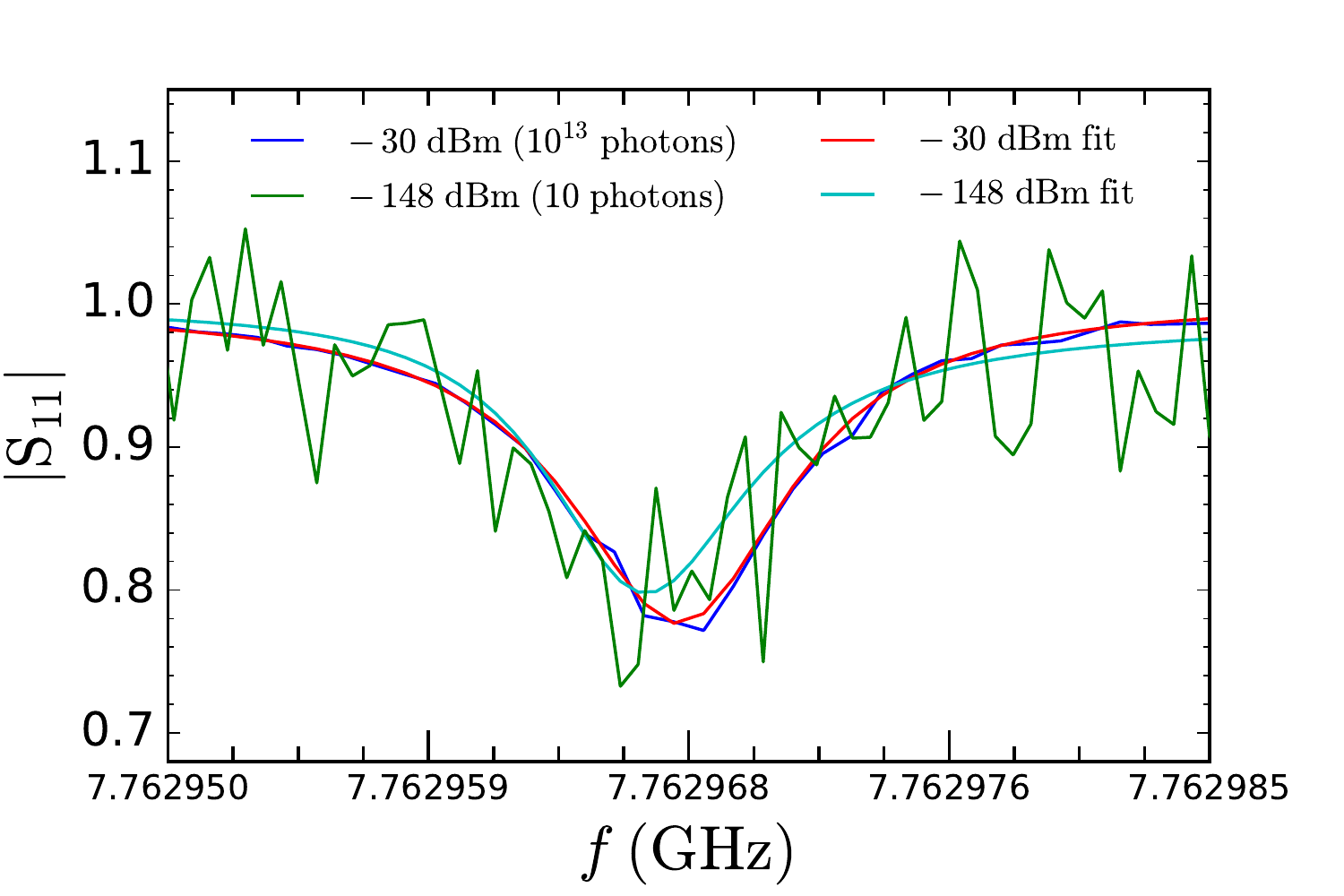}
	\caption{\label{fig3}  Graph overlaying reflection measurements at input powers of -30~dBm and -148~dBm, along with their fits. }
\end{figure}

\begin{figure}
	\includegraphics{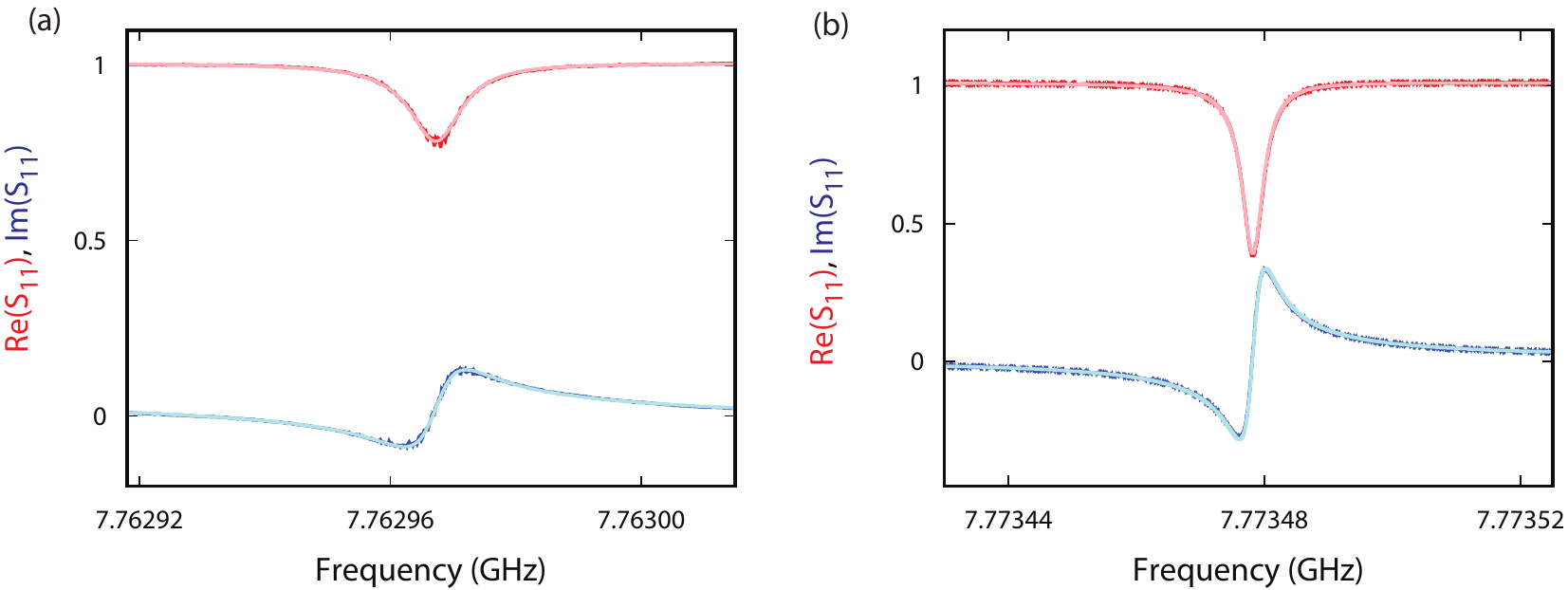}
	\caption{\label{fig4}  $S_{11}$ measurements of split cavities. Shown are the real and imaginary quadratures of the $S_{11}$ parameter of the aluminum cavity in (a) split geometry and (b) closed geometry. Derived values can be read from Table S1. Lighter lines indicate the fit. }
\end{figure}

\begin{table}[h]
	\begin{tabular}{|l|c|c|c|c|}
		Cavity state & $Q_\text{i}$ & $Q_\text{e}$ & $Q_\text{L}$ \\
		Split & $8.75 \pm 0.03 \times 10^5$ & $7.04 \pm 0.02 \times 10^6$ & $7.79 \pm 0.03 \times 10^5$ \\
		Closed & $2.748 \pm 0.001 \times 10^6$ & $6.242 \pm 0.001 \times 10^6$ & $1.908 \pm 0.001 \times 10^6$ \\
	\end{tabular}
	\caption{Quality factor measurements of our aluminium cavity in open and split configurations. The split is created by 100~$\mu$m Kapton tape. }
\end{table}
%\bibliography{}

\end{document}